\documentclass[usenatbib]{mn2e}
\usepackage{aas_macros,psfig,amsmath}

\input epsf

\begin{document}

\title[]{Large scale bias and
stochasticity of halos and dark matter}
\author[U.  Seljak \& M. S. Warren]{Uro\v{s} Seljak $^1$ \& Michael S. Warren $^2$\\
$^1$ Department of Physics, Jadwin Hall\\ Princeton University,
        Princeton, NJ 08544 \\
$^2$ Theoretical Astrophysics (T-6),
Mail Stop B227,
Los Alamos National Laboratory,
Los Alamos, NM 87545 }

\maketitle

\begin{abstract}
On large scales 
galaxies and their halos are
usually assumed to trace the dark matter with a constant bias
and dark matter is assumed to trace the linear density field.  
We test these assumption using 
several large N-body simulations with $384^3-1024^3$ particles and 
box sizes of $96-1152h^{-1}$Mpc, 
which can both resolve the small galactic size halos and 
sample the large scale fluctuations. 
We explore the average halo bias relation as a function of halo mass
and show that existing fitting formulae overestimate the halo bias by 
up to 20\% in the regime just below the nonlinear mass. 
We propose a new expression that fits our simulations well.
We find that the halo bias is nearly constant, $b \sim 0.65-0.7$, 
for masses below one tenth of the nonlinear mass.
We explore next
the relation between 
the initial and final dark matter  
in individual Fourier modes 
and show that there are significant 
fluctuations in their ratio,
ranging from 10\% rms at $k\sim 0.03$h/Mpc to
$50\%$ rms at $k\sim 0.1$h/Mpc. 
We argue that these large fluctuations are caused by perturbative 
effects beyond the linear theory,
which are 
dominated by long wavelength modes with large random fluctuations. 
Similar or larger fluctuations exist between halos and dark matter and 
between halos of different mass.
These fluctuations 
must be included in attempts to 
determine the relative bias of two populations from their maps, 
which would otherwise be immune to sampling variance. 

\end{abstract}

\section{Introduction}

Determination of the power spectrum of mass fluctuations and 
its redshift evolution is one of the 
main goals of modern observational cosmology. 
Its accurate measurement would allow us to test some of the most 
fundamental questions in cosmology today, such as the shape of 
primordial power spectrum and its relation to fundamental theories 
of structure formation, the mass of neutrino and the  
nature of dark energy. 

In general there are two approaches to the measurement of the matter 
power spectrum. One is to measure galaxies, 
either in redshift space or in angular position (perhaps supplemented by 
photometric redshift information) and to assume they trace the dark matter. 
This assumption is believed to be valid on large scales, where the so called
linear bias model assumes that the galaxy density 
field is proportional to the matter density field
times a free parameter called bias. While power spectrum measurements of
galaxies
with modern surveys such as SDSS \citep{2003astro.ph.10725T} or 2dF\citep{2002MNRAS.337.1068P}
have enormous statistical power, they can only determine the shape of the 
matter power spectrum and not its amplitude because of the bias uncertainties. 
This limits their use in the study of the growth factor evolution, important 
for investigations of dark energy models. In addition, 
on small scales information from 
galaxy clustering is limited by the uncertainties in the relation between 
the galaxies and the dark matter, which make the bias scale dependent. 
For this reason the small scale information is usually discarded.

The situation with galaxies would not be as dimm if we could 
determine the bias.
Here we explore a method to determine 
galaxy bias based on determination of clustering 
amplitude of faint galaxies. 
These are likely to occupy
low mass halos which, as we show in this paper,
have a well determined large scale bias that is nearly 
independent of halo mass. 
While some fraction of these galaxies are satellites in larger halos, 
this can be quantified and corrected for (see \cite{2004astro.ph..6594S} 
for a first application 
of this method to the real data). In addition, there exist 
populations, such as IRAS galaxies, for which this fraction may be small. 
In these cases
measuring the large scale power spectrum amplitude for these galaxies
determines the matter power spectrum amplitude as well. 

The main problem with using faint galaxies as tracers of large scale 
structure is that
in a typical flux limited survey 
faint galaxies occupy a small nearby volume, so the sampling variance 
errors for power spectrum on large scales are large. However, 
we can still 
determine their bias relative to a population of brighter galaxies. 
If there is no stochasticity between the two populations then a direct
comparison of the maps 
gives an accurate determination of the relative bias with no sampling 
variance. We can then use the power spectrum determination of the 
brighter population, with smaller sampling variance errors because of 
larger volume covered, to determine 
the power spectrum of the fainter population and of the matter itself.
The limiting source of noise is the stochasticity between these fields, 
which we explore in this paper. 

Another approach to determine matter fluctuations is to use weak 
lensing induced correlations between background galaxy ellipticities. These 
are sensitive to the dark matter fluctuations directly and 
as such this approach holds the promise to improve upon the limitations of 
the galaxy clustering methods. Its main limitation is that it 
traces the dark matter in angular projection and has large 
sampling variance errors on large scales. 
This limits the statistical 
power of the weak lensing surveys. 
On small scales the nonlinear
corrections, noise, intrinsic correlations and other systematic contaminations
become significant, all of which may complicate the modelling. 

A possible approach to achieve the best of both worlds 
is to combine the weak lensing and galaxy 
clustering surveys: one can use the weak lensing to determine the galaxy 
bias and then use the 3-d 
galaxy clustering information to improve on the statistical errors. 
One way to do this is to 
use galaxy-dark matter cross-correlation analysis from weak lensing 
and combine it with the 
galaxy auto-correlation analysis. 
If galaxies are tracing perfectly the dark matter then it suffices to 
have a few well measured modes in both fields to determine the galaxy bias. 
This has been proposed as a way to get around the sampling variance in weak 
lensing surveys \citep{2004astro.ph..2008P}. 
In the absence of stochasticity
it gives the bias (and so the dark matter power spectrum) without the 
usual sampling variance errors, assuming the analysis is done on the 
same patch of the sky and with the correct radial weighting of
the galaxies to match that of the dark matter.

In both of these cases the underlying assumption is that there is no stochasticity between 
these fields on large scales. 
While there have been analytic attempts to address this assumption
\citep{1999ApJ...525..543M}, 
it has not been tested well with simulations in
the past due to the lack of sufficient dynamic range (but 
see \cite{2002MNRAS.333..730C} for a related study). 
One must resolve the halos small enough to be suitable as galaxy 
hosts (with typical masses at or below $10^{12}h^{-1}M_{\sun}$). At the 
same time, the simulations must be large enough so that many long wavelength
modes are sampled to determine the statistics of interest. 
We achieve this by using a set of new simulations with a larger dynamical range. 
The number of particles in these simulations, 
$10^8-10^{9}$, and their box size, 100-1000$h^{-1}$Mpc, allow a much 
better exploration of the halo bias and stochasticity on scales larger
than available before. 

In addition to exploring the relation between halos and matter 
we can also investigate the relation between the initial and final 
matter distribution.
Weak lensing measures the
nonlinear matter field, while 
for the study of linear growth factor one would like to know 
the relation between galaxies and linear matter field instead. 
We explore the relation between 
the final and initial dark matter density 
field on 
large scales, where this relation is believed to be perfect. 
This case is amenable to perturbation theory analysis 
and as such allows one to interpret and verify
the numerical simulation results. 

Finally, we revisit the question of halo bias as a function of halo mass with the 
new simulations. 
This
has been addressed by previous generation of simulations using 
$256^3$ particles and box sizes of order (100-140)$h^{-1}$Mpc
\citep{1998ApJ...503L...9J,1999MNRAS.308..119S,2001MNRAS.323....1S}.
This, as noted by the authors themselves, is 
barely adequate for this purpose because of large shot noise at high halo masses 
and insufficient number of large scale modes where linear evolution is valid. 
The goal of these papers was to provide expressions
which fit over a range of power law simulations and were not specifically 
optimized for realistic $\Lambda $CDM models. 
They 
provide expressions that fit the simulations available at the time 
to a reasonable accuracy, but which can be systematically wrong by 
as much as 10-20\%. 
To put things into a 
current context, the statistical error on the amplitude of galaxy clustering 
from SDSS using $k<0.2h$/Mpc modes is 1\% \citep{2003astro.ph.10725T}, so a perfect bias 
determination, for example using the faint galaxies as described above,
would allow us
to reach this accuracy on the matter power spectrum. In this era 20\%
accuracy no longer suffices and the goal of the present paper is to 
provide more accurate expressions for halo bias as a function of mass 
and cosmological model. 

\section{Simulations}

The N-body code we use in this paper is the
Hashed Oct-tree code (HOT), a parallel, tree based, code \citep{WarSal93a}.
This code was compared to a variety of other simulation codes
	in \cite{1999ApJ...525..554F}, and further validation studies will be
	presented elsewhere. 
For this paper we performed several simulations with this code. The smallest 
was a 96$h^{-1}$Mpc box size, $512^3$ particle run (HOT1). 
This simulation has a particle mass of $5.5\times 10^8h^{-1}M_{\sun}$ and 
is useful for probing the halo bias at the low mass end, below $10^{11}h^{-1}M_{\sun}$. It
suffers from the small box size which makes the investigations on linear 
scales difficult and makes the shot noise fluctuations for higher mass halos 
(with lower halo numbers) very large. 
Next up in size is a simulation with 288$h^{-1}$Mpc box and $768^3$ particles (HOT2). This is 
the main simulation that we use in this paper, as it has an optimal combination 
of box size and particle mass for our purposes. It samples the Fourier modes down 
to $k \sim 0.02h$/Mpc and has many modes at $k \sim 0.1h$/Mpc, where the power 
spectrum is still close to linear. This is also 
the typical scale probed by the
current surveys such as SDSS and 2dF. 
The particle mass for this simulation 
is $4.4\times 10^9h^{-1}M_{\sun}$ and can resolve halos down to 
a few times $10^{11}h^{-1}M_{\sun}$, which is sufficient for typical  
galaxies in a flux limited sample.
Finally, for determination of 
the halo bias at the high mass end we use a simulation with 
1152$h^{-1}$Mpc box size and $768^3$ particles (HOT3). This simulation has large enough box
to sample long wavelength modes well, but its particle mass of 
$2.8\times 10^{11}h^{-1}M_{\sun}$ 
does not allow us to resolve 
galactic size halos and we limit its use to group and cluster size halos. 

	The tree-code accuracy was controlled using the absolute error
	criterion described in \cite{SalWar94a}, which ranged from 
	$10^{-5} M_{tot}/R_0^2$ per interaction
	at the start of the each simulation, to $10^{-3} M_{tot}/R_0^2$ at
	the end.
	Plummer smoothing was used, with softening lengths of 
	7, 20 and 95 comoving kpc for models 1, 2 and 3 respectively.
	The number of timesteps for model 1 was 1475, 736 for model 2,
	and 725 for model 3.
	Model 1 started at a redshift of 50, model 2 at 44, and model
	3 at 27.
	All particle masses were identical, with the initial particle
	displacements imposed on a cubical lattice.
	
All of these simulations are normalized to $\sigma_8=0.9$ and have $\Omega_m=0.3$, $\Omega_b=0.04$ and Hubble parameter $h=0.7$. They 
use realistic transfer functions from CMBFAST \citep{1996ApJ...469..437S}. 
$\sigma_8 = 0.9$ corresponds to $\delta_\zeta = 4.624 \times 10^{-5}$ 
normalization in CMBFAST.
In the following all the results are for HOT2 
whenever not explicitly specified otherwise. 

For the purpose of studying bias as a function of halo mass we also 
ran a suite of simulations varying one parameter at a time. The
box size for these simulations is 192$h^{-1}$Mpc with $512^3$ particles. Their 
force and mass resolution is the same as for HOT2. We ran the basic 
simulation with the same parameters as for HOT1-3, as well as 
$\Omega_m=0.2$, $\sigma_8=0.8$, $n=0.9$, $dn /d \ln k=-0.04$ and $h=0.6$, 
6 simulations in total. We used the same seed in random generator for all of 
these cases to minimize the sampling errors. To investigate the bias 
at low mass end we supplemented these simulations with another 
run with $512^3$ particles in 96$h^{-1}$Mpc. 

Finally, to investigate the bias at the high mass end we used additional 
simulations with $384^3$ particles in 768$h^{-1}$Mpc box, with standard 
parameters and $\sigma_8=0.90$, $\sigma_8=0.775$ and $\sigma_8=1.046$. We also used another
simulation with 700$h^{-1}$Mpc and $512^3$ for which 
$\sigma_8=0.767$, $\Omega_m=0.27$ and $h=0.71$. While this paper
was undergoing the refereeing process we finished  another
very large
simulation with 768$h^{-1}$Mpc box and $1024^3$ particles, again with 
the standard parameters. We use this simulation to verify the results
obtained with other simulations, finding a good agreement among them.

\section{Stochasticity of dark matter}

\begin{figure}
\centerline{\psfig{file=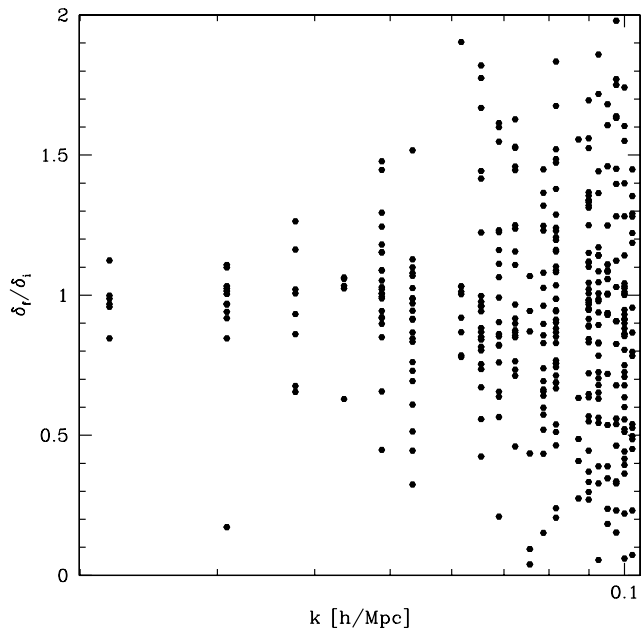,width=3.5in}}
\caption{Ratio of final to initial density perturbations as a function of 
wavemode amplitude $k$. There is a large scatter between the two quantities even on 
large scales, where linear theory is usually assumed to be valid. 
}
\label{fig1}
\end{figure}

We begin by exploring the relation between the final dark matter 
density field and the initial density field, rescaled to $z=0$ using 
the linear growth factor. We Fourier 
transform both fields and denote individual modes with $\delta_i(\bf{k})$ 
and $\delta_f(\bf{k})$ (we 
treat real and imaginary components as separate modes).
Figure \ref{fig1} shows 
the ratio $b(\bf{k})=\delta_f(\bf{k})/\delta_i(\bf{k})$ 
for $k<0.1$h/Mpc. This is the scale at which one often assumes
linear theory to be valid.
We see that there are significant fluctuations between the initial and final 
field, suggesting that there are large corrections to the 
linear evolution even for $k<0.1$h/Mpc. 

\begin{figure}
\centerline{\psfig{file=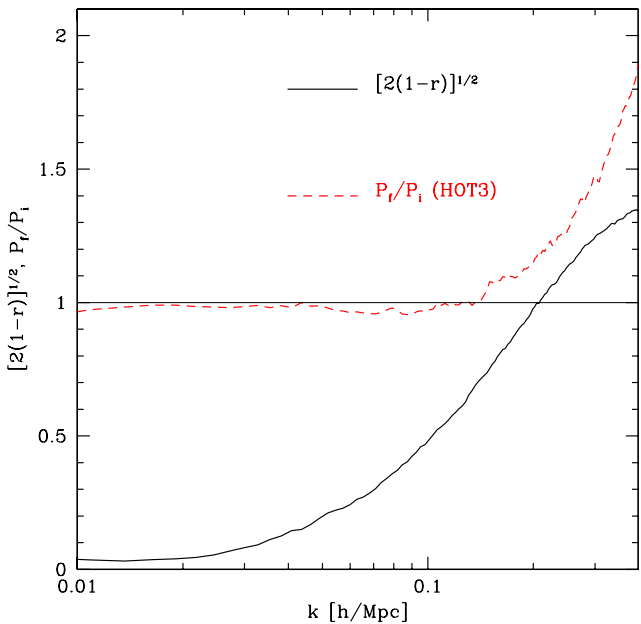,width=3.5in}}
\caption{Relative rms fluctuation $\sigma_b/b=[2(1-r)]^{1/2}$ between the 
final and the
initial density field for HOT3 (solid). Also shown is the power spectrum ratio
between the two fields (dashed). Results for HOT2 are similar, but show a 
somewhat larger supression of nonlinear versus linear power spectrum for 
$k<0.1h/Mpc$. 
}
\label{fig2}
\end{figure}

We can define the ratio of the power spectra as
\begin{equation}
\langle b^2(k) \rangle \equiv {P_f(k) \over P_i(k)}
={\langle \delta_f^2(\bf{k}) \rangle \over \langle \delta_i^2(\bf{k})
\rangle},
\end{equation}
where $\langle \rangle$ denotes ensemble average over different realizations
of the universe.
We can define relative rms fluctuations in 
$b$ as  
\begin{equation}
\left({\sigma_b \over b}\right)^2={\langle (\delta_f -\langle b^2 \rangle^{1/2}\delta_i)^2 \rangle \over 
\langle \delta_f^2 \rangle}.
\end{equation}
This is related to 
the cross-correlation coefficient $r$, defined as 
\begin{equation}
r(k)={\langle \delta_i(\bf{k})\delta_f(\bf{k}) \rangle \over 
\sqrt{\langle \delta_i(\bf{k})\delta_i(\bf{k}) \rangle\langle \delta_f(\bf{k})\delta_f(\bf{k}) \rangle}},
\label{r}.
\end{equation}
The two are related via
\begin{equation}
{\sigma_b \over b}=\sqrt{2(1-r)}.
\end{equation}
Figure 
\ref{fig2} shows 
$\sigma_b /b$ as a function of wavemode $k$, 
where the average has been done over a large number of wavemodes so that 
$r$ converges (at very low $k$ this condition is not satisfied and $r$ is biased high, 
which underestimates the rms fluctuations).
We see that $\sigma_b / b$ changes from 10\% at $k \sim 0.02h$/Mpc 
to 40\% at 
$0.1$h/Mpc, above which it rapidly increases and the two fields become incoherent.
Figure \ref{fig2} also shows the ratio of nonlinear to linear power 
spectrum at $z=0$, $\langle b^2(k) \rangle$. 
For $k>0.15$h/Mpc the final power spectrum 
rapidly grows with $k$ and exceeds the linear power spectrum,
while for $k<0.15$h/Mpc 
the final spectrum is slightly anti-biased on large scales, ie $P_f(k)<P_i(k)$. 
We note that this effect is larger when small boxes are used. We discuss 
this further below. 

The main result arising from figures \ref{fig1}-\ref{fig2} is that the fluctuations between 
the linear and nonlinear 
fields are large on large scales, despite the fact that the nonlinear power 
spectrum is very close to the linear one. This is not so evident 
from the cross-correlation coefficient $r$ itself, which can be close to 
1 and still lead to large rms fluctuations: 
even for $r=0.995$ the rms 
fluctuations between initial and final field are 10\% 
for any given mode. 

One can get some understanding of these
results by using second order perturbation theory results 
\citep[see][ for a recent review]{2002PhR...367....1B}. 
To compute the power spectrum to second order one must derive the 
density field to 3rd order, $\delta=\delta_1+\delta_2+\delta_3$. 
Second order contributions to the power spectrum 
arise  both from $\delta_2\delta_2$ and 
$\delta_1\delta_3$ terms. The $\delta_2\delta_2$ is the mode-mode coupling term, 
while the $\delta_1\delta_3$ is the nonlinear growth evolution term. These terms 
have different behaviour in various limits and have differing
signs in the contribution: while $\delta_2\delta_2$ is strictly positive, 
$\delta_1\delta_3$ has a negative component. 
For $k<0.15h$/Mpc the negative contribution wins 
and the second order 
correction to the power spectrum is negative. At its peak around 
$k \sim 0.1h$/Mpc the correction is 5\% and  slowly decreases towards
$k \rightarrow 0$. While the correction is small on average, this is a 
result of a cancellation between positive and negative contributions.
The dominant perturbative corrections come from the mode-mode couplings
at wavelengths
close to the wavelength of the mode itself: 
for $k=0.1$/Mpc the dominant contribution to 
the positive component is from the modes around $k\sim 0.05$/Mpc, which 
contribute around a third of the total correction or 
around 5\% of the final power spectrum \citep{1994ApJ...431..495J}. These are long wavelength 
modes and in any finite volume there will be large statistical fluctuations 
in their power relative to the true power. 
This leads to significant fluctuations in the second order corrections depending 
on the actual realization of the mode amplitudes. Thus the
final amplitudes of individual modes fluctuate significantly 
relative to their initial values because of perturbative 
large scale effects. 

A similar effect is observed in the average nonlinear power spectrum, which 
is suppressed relative to linear one. 
The suppression of nonlinear 
power for $k<0.15h/Mpc$ relative to linear one is dominated by long wavelength 
mode-mode coupling, so to get to a percent precision in simulations 
one needs very large simulation boxes. We find that the difference between 
HOT2 (320$h^{-1}$Mpc) and HOT3 (1152$h^{-1}$Mpc) is 5\% in power 
at k=0.1h${-1}$Mpc, 
with larger HOT3 simulation 
being closer to the linear power spectrum than HOT2. 
Thus while these mode-mode induced 
fluctuations are small compared to sampling variance 
for individual modes, they are not small when one averages
over many modes and may dominate the accuracy of amplitude determination 
on large scales. 

\section{Stochasticity of halos}

Galaxies are believed to form inside dark matter halos, which are virialized 
structures of high density. They can be labelled by their virial mass. 
Observations suggest that about 80\% of the galaxies in a typical flux 
limited survey form at the 
centers of halos with masses ranging between $10^{11}h^{-1}M_{\sun}$ to 
$10^{13}h^{-1}M_{\sun}$, while the remaining 20\% of galaxies are non-central and occupy 
groups and clusters \citep{2002MNRAS.335..311G,2004astro.ph..6594S}. 
The exact radial distribution of galaxies inside halos 
and the form of the halo mass probability distribution
is the subject of a lot of current observational and theoretical effort. 
Here we will use centers of dark matter halos as a proxy for galaxy positions. 
This will not give the correct correlation properties 
on small scales, where correlations between central and noncentral galaxies
within the halos are important, but should be valid on large scales, where halos 
can be thought of as pointlike. We will show the results for a range of halo masses, 
which can be roughly thought 
as corresponding to galaxies with different luminosities 
since there is a tight
relation between the halo mass and luminosity 
\citep{2001astro.ph..8013M,2002MNRAS.335..311G,2004astro.ph..6594S}. 
Alternatively, the different samples can be thought of as varying the 
flux limit of a survey, since going to fainter limits increases the 
number density of galaxies and thus reduces the shot noise and the same
effect is achieved by going to fainter galaxies. 

\begin{figure}
\centerline{\psfig{file=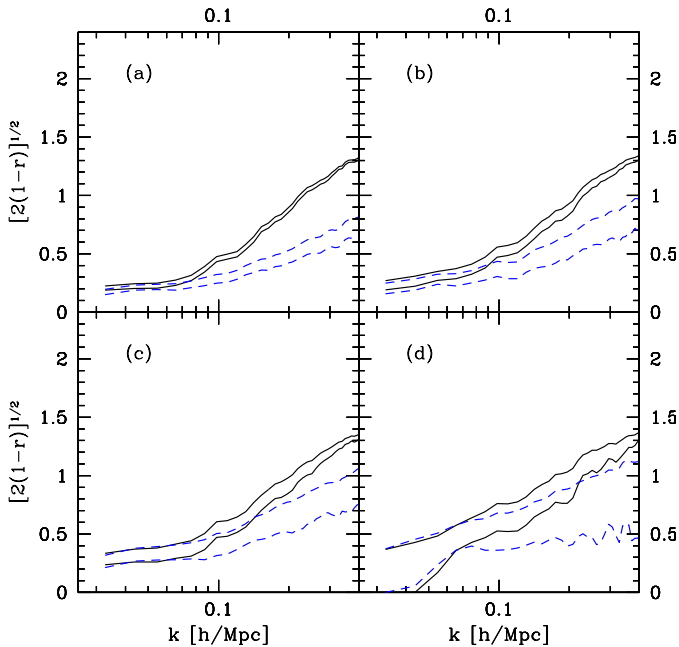,width=3.5in}}
\caption{Relative rms fluctuation $\sigma_b/b=\sqrt{2(1-r)}$ between the halo density 
field and the initial (solid) or final (dashed) matter density field.
Lower curves have been obtained by applying 
the shot noise subtraction from the halo power spectrum. Average masses are 
$4.5\times 10^{11}h^{-1}M_{\sun}$ (a), $10^{12}h^{-1}M_{\sun}$ (b), 
$2\times 10^{12}h^{-1}M_{\sun}$ (c) and $10^{13}h^{-1}M_{\sun}$ (d).
The corresponding halo densities are 
$7\times 10^{-3}h^3/Mpc^3$, $2.7\times 10^{-3}h^3/Mpc^3$, $1.5\times 10^{-3}h^3/Mpc^3$
and $3.5\times 10^{-4}h^3/Mpc^3$.
}
\label{fig3}
\end{figure}

Dark matter halos are identified from the simulations using the standard 
friends of friends algorithm with a linking length of 0.2. The resulting 
mass functions agree well with the fitting formulae in the 
literature \citep{1999MNRAS.308..119S,2001MNRAS.321..372J}. 
We order them by mass and use subsamples separated roughly by a factor of 2 in mass. 
As in previous section we can define the halo fluctuation $\delta_h(\bf{k})$
and bias $b({\bf k})=\delta_h(\bf{k})/\delta_m(\bf{k})$, as well as the 
cross-correlation coefficient between the two fields (equation \ref{r}). 
Figure \ref{fig3} shows the relative rms fluctuations $\sigma_b/b$ as a function 
of scale for several halo masses, relative to both the initial and the final 
density 
field. 
We show the case with and without the subtraction of
shot noise contribution to the
halo power spectrum (the dark matter power spectrum does not require shot noise 
subtraction because
of large number of dark matter particles). 
The lines without shot noise subtraction are always above the 
ones with subtraction and are the relevant ones if one is interested in the 
stochasticity between the halos and dark matter. The lower lines for which 
the shot noise has been subtracted show the remaining stochasticity which is 
not due to the shot noise. 
Because of the shot noise subtraction 
the cross-correlation coefficient can exceed 1, in which case we do not show the result. 

From figure \ref{fig3} we see that the halos are even 
less well correlated to the
initial density field than the dark matter is.
The stochasticity begins at the level of 20\%
at low $k$, increasing to 50\% at $k\sim 0.1h$/Mpc. The shot noise contribution 
to stochasticity 
is small for halos with high spatial density (low mass halos), but increases significantly 
for halos with low spatial density, as expected. There is no obvious difference in 
the shot noise subtracted values, 
suggesting that the shot noise simply adds
an additional component of stochasticity on top of that induced by nonlinearities in 
the relation between the halos and the initial density field. 

Correlation coefficient between the
halos and the final density field is also  shown in figure \ref{fig3} (dashed lines).  
Compared to the halo-initial field correlations 
the stochasticity is similar on the largest scales, but 
 there is a better agreement between the halo and the final dark matter 
field on smaller scales ($k>0.1h$/Mpc). 
The cross-correlation coefficient $r$ would 
likely be even larger on small scales if we had modelled the galaxy distributions
within the halos more realistically, since this would lead to an 
enhancement of correlations on small scales, similar to that seen 
in the dark matter. 
Results from GIF simulations and analytic results using halo models suggest 
that the cross-correlation coefficient can remain close to unity 
up to a fairly high $k \sim 1$h/Mpc \citep{2000MNRAS.318..203S}, 
but this may not be generic and depends on the details 
of how galaxies are populated within the halos, which are quite
uncertain. 
Observational evidence suggests that there is some stochasticity 
on 1Mpc scale, with $r\sim 0.5$ \citep{2002ApJ...577..604H}.
If $r<1$ it would complicate the interpretation of the results based on the
comparison between galaxy-galaxy correlations and
galaxy-dark matter correlations, such as those from the galaxy-galaxy lensing
analysis \citep{2003astro.ph.12036S}. 
Here we are more concerned with the correlations on large scales, $k<0.1$h/Mpc, 
where the details of galaxy distribution within halos are not important 
and where direct observations are not yet available. 
The results suggest that the fluctuations between halos and initial or 
final matter field are never below 10-20\%.  

The dark matter distribution cannot be directly 
observed, so results shown in figure \ref{fig3} are not directly 
applicable to any 
observational test. The closest example 
to a direct observation of the dark matter 
is through the weak lensing effect. Here the light from distant galactic sources
is being distorted by the mass distribution along the line of sight. 
By averaging over the image distortions we can reconstruct the 2-d shear and 
convergence maps. 
These are given by the line of sight projection of the matter density, weighted
by a radial function that is very broad. Correlations at a given angular scale 
receive dominant contributions from a transverse distance at half the distance to 
the source, but significant contributions are also coming from much smaller 
transverse separations produced by the mass distribution closer to the observer. 

It has been suggested by \cite{2004astro.ph..2008P} that if one cross-correlates the properly 
radially weighted galaxy field with the weak lensing maps then one determines
the bias of galaxies exactly if the two are perfectly correlated. Under 
these assumptions one can use the galaxy clustering information to determine 
the amplitude of dark matter fluctuations with higher accuracy than from the 
weak lensing itself, because the 
galaxy clustering can be done in 3-d (if redshifts
are measured) and so one has more independent modes to reduce the sampling 
variance compared to the 2-d analysis. For this method to work the correlation 
between the projected matter density and galaxy field must be close to perfect.

\begin{figure}
\centerline{\psfig{file=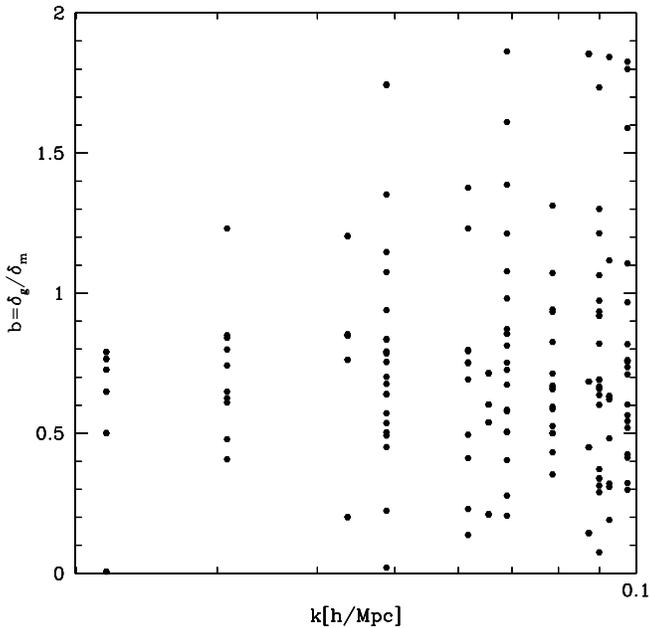,width=3.5in}}
\caption{
Ratio of 2-d projected halo density perturbations ($M=10^{11}h^{-1}M_{\sun}$)
to the initial density field as a function of
wavemode amplitude $k$. The projections are along each of the three axes
(288$h^{-1}$Mpc for HOT2 simulation used here). 
The scatter is larger than the corresponding 3-d case in figure 
\ref{fig3}. 
}
\label{fig5}
\end{figure}

To address this assumption one must correlate 2-d projections of final dark 
matter and galaxies. 
While properly projected weak lensing 
2-d maps have been constructed from N-body simulations \citep{2000ApJ...530..547J,2000ApJ...537....1W}, 
we take a simplifying approach here and cross-correlate the 2-d projections
of the simulations along each of the 3 axes. The resulting rms scatter as a function 
of projected wavevector $k$ is found to be significantly larger than in 
3-d 
case, a consequence of the projection effects, which cause shorter 
wavelength modes
to contribute to longer wavelength modes in projection. 
For $10^{12}h^{-1}M_{\sun}$
halos, which corresponds roughly to $L_*$ galaxies,
we find that the rms scatter is 20\% at $k\sim 0.03h$/Mpc and 
40\% at $k\sim 0.1h$/Mpc. This is reduced by a factor of 2 if 
$10^{11}h^{-1}M_{\sun}$ galaxies are used instead.
This last example is shown in figure \ref{fig5}. 
In reality the stochasticity is 
likely to be larger for the lensing case, 
since projections at a fixed angle (rather than at a fixed transverse 
separation as done here) receive contributions from nearby small 
scale structures, for which the stochasticity between the galaxies and the dark matter 
will be much larger.

We can estimate the effect of this scatter on the amplitude determination 
from the weak lensing cross-correlation analysis. 
If the lensing 
kernel peaks at $z \sim 0.3-0.4$ then $k \sim 0.1$h/Mpc corresponds to $l\sim 100$. 
In a 200 square degree survey such as the 
upcoming CFHT Legacy Survey \citep{2003astro.ph..5089V}
we will have about 5 independent modes at $l\sim 30$ and 50 at $l \sim 100$. 
This means that for galaxies in $10^{12}h^{-1}M_{\sun}$ halos 
the overall linear amplitude will have an error  
of 20\%/$\sqrt{5}\sim $ 
9\% at $l \sim 30$ and 6\% at $l \sim 100$, arising just from this effect 
(the power spectrum amplitude error will be twice as large). 
Additional errors of comparable magnitude will arise from the lensing noise and projection
effects. Such a 
poor determination of the growth factor as a function of redshift
is unlikely to 
improve our current constraints on the dark energy significantly.
This source of 
error was not included in the previous analysis \citep{2004astro.ph..2008P} and 
is much larger than the prognosticed errors without it. This
complicates the prospects of 
this method for studies of dark energy through the
growth factor evolution.
The errors can be reduced with a larger survey area: 
for a survey covering 25\% of full sky the errors on the power spectrum 
amplitude may approach 1\% because more modes are being sampled and 
because the largest modes have the smallest amount of stochasticity. 
It remains to be seen whether this is ever competitive with 
a straight weak lensing analysis on smaller scales.

\begin{figure}
\centerline{\psfig{file=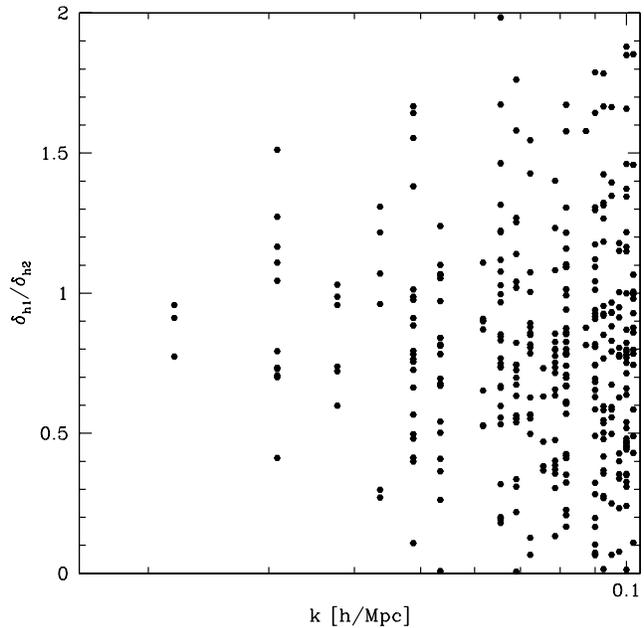,width=3.5in}}
\caption{
Ratio of halo density perturbations $\delta_{h1}/\delta_{h2}$ as a function of
wavemode amplitude $k$. The halos are of mass
$10^{11}h^{-1}M_{\sun}$ (h1) and $10^{12}h^{-1}M_{\sun}$ (h2). 
}
\label{fig4}
\end{figure}

As discussed in the introduction another method to determine the 
bias is to combine the clustering analysis of faint galaxies, for 
which we know the theoretical bias, with the luminous galaxies, for 
which we can measure the clustering on large scales with a small 
statistical error. To determine the relative bias between the populations
we can simply compare the smoothed maps. In the absence of stochasticity 
between the two galaxy populations
one could determine 
the amplitude of mass fluctuations directly.
Suppose that we want to determine the clustering 
amplitude of faint galaxies, which 
are in low mass halos (around $10^{11}h^{-1}M_{\sun}$ for galaxies 
2-3 magnitudes below $L_*$) and $L_*$ galaxies, which typically 
occupy $10^{12}h^{-1}M_{\sun}$. 
Figure \ref{fig4} shows the relative rms fluctuations in 
ratios of Fourier mode amplitudes
between halos of mass $10^{11}h^{-1}M_{\sun}$ and $10^{12}h^{-1}M_{\sun}$. 
We see again from figure \ref{fig4} that the scatter is large. 
Both shot noise
and stochasticity due to nonlinearities limit this method. 
The rms fluctuations between $10^{11}h^{-1}M_{\sun}$ and 
$10^{12}h^{-1}M_{\sun}$ halos are 8\% at $k \sim 0.1h$/Mpc
and 23\% at $k \sim 0.2h$/Mpc.
This is somewhat smaller than between the
halos and the matter. Moreover, galaxies in redshift surveys
provide 3-d information, so there are more large scale modes to 
reduce the scatter.
Nevertheless,
any attempt to determine the linear bias using the 
cross-correlations must include this source of stochasticity in the analysis. 

\section{Halo bias as a function of mass}

One of the important questions that can be addressed with these
simulations is the relation between halo and dark matter power spectrum as a 
function of halo mass. 
This relation has been theoretically predicted from the spherical collapse
model \citep{1989MNRAS.237.1127C,1996MNRAS.282..347M} and from the ellipsoidal collapse model 
\citep{2001MNRAS.323....1S}, which suggest that the halo bias is related to a derivative of the 
halo mass function. The relation
has also been extracted from the numerical simulations, with a good quantitative 
agreement
between the theoretical predictions and the simulations over a range of different
cosmological models \citep{1998ApJ...503L...9J,1999MNRAS.308..119S}. 
Since these comparisons were done for a wide range 
of initial power spectra they were not specifically designed for 
realistic $\Lambda$CDM models 
and the predictions and simulations could differ by up to 20\%.  
Moreover, the simulations used in previous work were based on $256^3$ 
particle simulations and did not have sufficient dynamic range to 
sample the long wavelength modes and resolve small halos at the same time. 
For the more massive halos shot noise is large,
so the bias estimate is noisy. For halos close to the resolution
threshold (typically of the order of 50-100 particles) some fraction of the
halos may be missed by the halo finder, leading to biased results in 
the bias determination as the low mass end.

The simulations used in this paper are a significant improvement over the
previous generation. They contain 8-64 times more 
particles and cover a wide range of masses. 
We use HOT1 simulation for halos in the mass range 
$(5\times 10^{10}-3\times 10^{11})M_{\sun}$,
HOT2 for halos
in the mass range  $(3\times 10^{11}-10^{14})M_{\sun}$ and
HOT3 for halos in the mass range 
$ (10^{13}-10^{15})M_{\sun}$ (the latter two are also 
checked with 768$h^{-1}$Mpc box $1024^3$ particle simulation). 

\begin{figure}
\centerline{\psfig{file=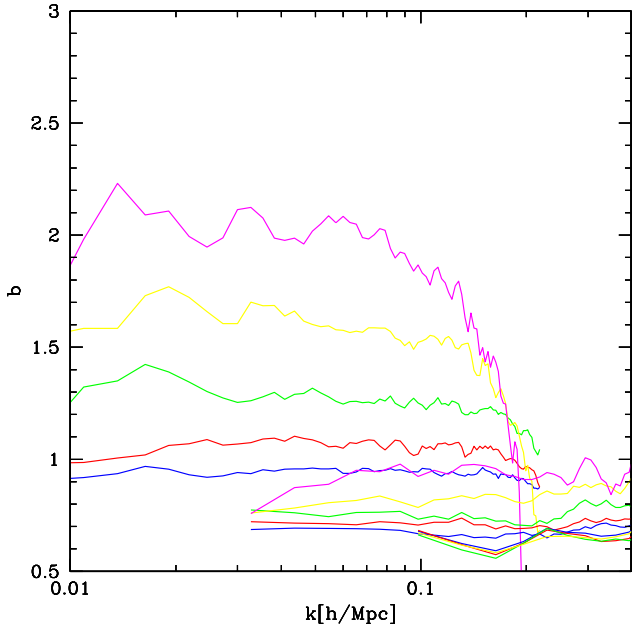,width=3.5in}}
\caption{Ratio of halo to linear density field power spectrum 
as a function of wavevector $k$ 
for halos of varying mass. 
At the bottom are the halos from HOT1 simulation, next up are those
from HOT2 and at the top are the HOT3 halos. 
}
\label{fig6}
\end{figure}

Figure \ref{fig6} shows the ratio of the (shot noise corrected) 
halo power spectrum to the linear mass 
power spectrum as a function of a wavevector $k$. One can see that the assumption 
of constant bias is reasonable for $k<0.1$h/Mpc and even beyond, 
so a linear bias can be 
defined as an appropriate 
average over these modes. The exception are the most massive halos
in HOT3 with $b>1.5$, for which the power spectrum is suppressed already 
at $k \sim 0.1$h/Mpc
due to the fact that the FOF halos do not overlap and so 
cannot be closer than two times the virial radius.
Here we use all of the modes with $k<0.1$h/Mpc,
except for the smallest 96$h^{-1}$Mpc simulation where we use $k<0.15$h/Mpc.
We note that there is a good agreement between the simulations in the 
overlap mass range, but the larger simulation 
has smaller statistical errors. The smallest simulation (HOT1) has very few
modes in the linear regime and the fluctuations in the ratio 
caused by perturbative effects beyond linear theory are large, 
so the bias determination from this simulation is somewhat less reliable. 
On the other hand, all of the low mass halos in this simulation have almost 
the same bias and at the upper end of the mass range there is a good 
agreement in bias with halos of the same mass from HOT2.

\begin{figure}
\centerline{\psfig{file=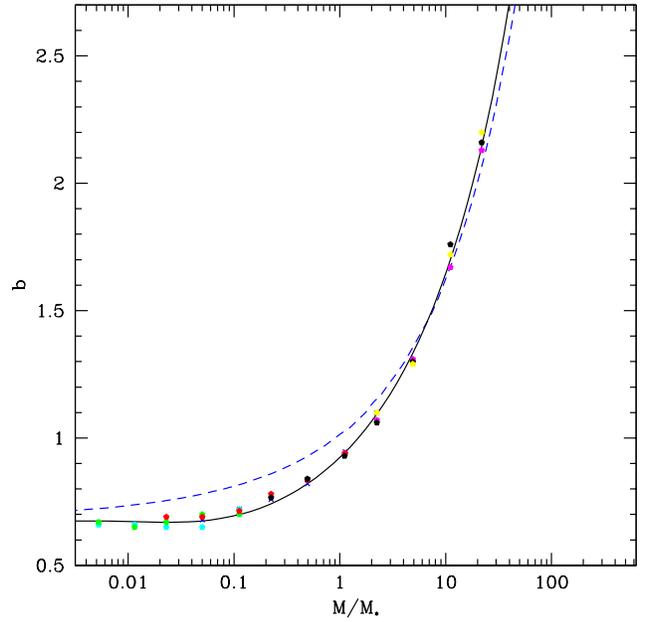,width=3.5in}}
\caption{Bias as a function of mass in units of the nonlinear mass. 
Points are from 96$h^{-1}$Mpc $512^3$ (HOT1, green), 
144$h^{-1}$Mpc $512^3$ (cyan), 192$h^{-1}$Mpc $512^3$ (red), 
288$h^{-1}$Mpc $768^3$ (HOT2, blue),  
1152$h^{-1}$Mpc $768^3$
(HOT3, magenta) and 
768$h^{-1}$Mpc $1024^3$ (black) simulations.  
Note that in several cases the points from 
two simulations overlap exactly. Upper (dashed blue) line is theoretical prediction 
from Sheth and Tormen (1999). 
Lower (solid black) line is the expression from
equation \ref{bfit}. 
}
\label{fig7}
\end{figure}

For simplification with theoretical comparisons we will
scale all the masses relative to the nonlinear mass $M_{\rm nl}$, defined
as the mass within a sphere for which the rms fluctuation amplitude of the linear 
field is 1.68. 
While the theoretical 
predictions for the bias depend on the cosmological model, most of that dependence
is accounted for if the mass is expressed in terms of the nonlinear mass. 
For HOT1-3 simulations with $\sigma_8=0.9$ and $\Omega_m=0.3$ at $z=0$
the nonlinear mass defined with 1.68 overdensity 
is $8.73\times 10^{12}h^{-1}M_{\sun}$. 
Figure \ref{fig7} shows the bias determinations as a function of halo mass
from the simulations used in this paper. 
The dashed line is the 
theoretical prediction from \cite{1999MNRAS.308..119S} (the fitting formula
given in \cite{1998ApJ...503L...9J} is very similar; while these 
fitting formulae are not very accurate 
we find a good agreement between the simulation results in these papers and 
our simulations). 
We see that these theoretical predictions overestimate
the bias below $M_{\rm nl}$ and are a good fit above $M_{\rm nl}$. 
The largest discrepancy is below $M_{\rm nl}$, where the relative error can be 
up to 20\%. The various simulations are in a reasonable 
agreement among themselves and 
the scatter between the points at 
the same mass is mostly due to the shot 
noise and small volume over which 
one is averaging. 
There may be some systematical error due to the fact 
that the nonlinear mass computed from the theoretical power spectrum 
can differ from the value obtained if one uses the actual realization. 
We find this can lead up to a 10\% effect on nonlinear mass and would cause
a horizontal shift by this amount. This is of almost  
no consequence for masses below $M_{\rm nl}$,
where bias is only weakly dependent on the mass, but may lead to a 
larger error at the high mass end. 

\begin{figure}
\centerline{\psfig{file=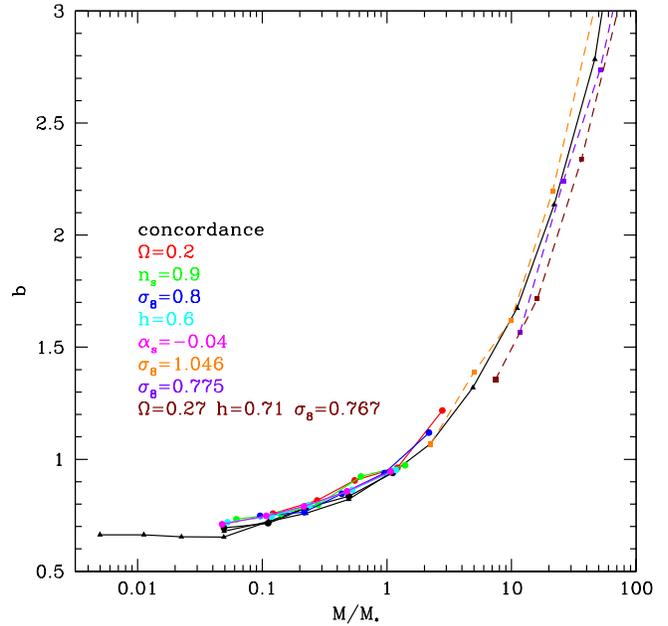,width=3.5in}}
\caption{Bias as a function of mass in units of the nonlinear mass
for several cosmological models. We varied one parameter at a 
time relative to the fiducial concordance model, roughly covering
the range of interest.
This figure shows that the bias predictions depend predominantly on
the nonlinear mass, while other cosmological parameters play only
a minor, but not entirely negligible, role.
}
\label{fig8}
\end{figure}

We find that the unbiased galaxies with $b=1$
are at $M=1.5M_{\rm nl}$ and the 
bias is rapidly changing above $0.1M_{\rm nl}$, while below this it is 
essentially constant with the value around 0.68. In all 
simulations we see
bias increasing at the lowest masses  
(figure \ref{fig8}), which is a numerical artifact. For example,  
such an increase is seen in HOT2 at the 
low mass end and is not confirmed in HOT1, where
the mass resolution improves by a factor of 8 (figure \ref{fig8}).
Moreover, this increase at low mass end 
changes into a decrease if we remove unbound 
particles from the halos. To be safe we only 
present results where the difference
between the two cases is less than 0.01. Note that in HOT1 we find
$b \sim 0.65$ at the low mass end. Even in the region of overlap 
with HOT2 the bias in HOT1 is systematically lower by 0.03. 
This is likely to be 
due to the sampling variance in HOT1, as can be seen from figure \ref{fig6}, 
which shows considerable fluctuations as a function of wavevector for 
this simulation. 
With a 144$h^{-1}$Mpc box $512^3$ particle simulation we again find that 
$b \sim 0.65-0.68$ at the low mass end and that 
there is indeed significant scatter due to small box size (figure 
\ref{fig7}). 
For this reason
the empirical fit given below goes above HOT1 at the low mass end. 
It is not entirely clear that this is the correct procedure, as HOT2 
could have been already affected by the resolution, but the fact that 
both unbound and bound halos give the same result argues against this. 

While there is some uncertainty in the bias value at the low mass end, 
all simulations agree very well around nonlinear mass. In addition 
to HOT2 simulation we also have another $512^3$ simulation with 192$h^{-1}$Mpc 
box simulation and a 768$h^{-1}$Mpc box with $1024^3$ particles 
simulation
that both sample well this regime. At the high mass
end uncertainty increases again because of a small number of high mass 
halos. In addition to HOT3 and 768$h^{-1}$Mpc box with $1024^3$ particles
simulation
we use another 768$h^{-1}$Mpc simulation with 
$384^3$ particles. 

The solid curve in figure \ref{fig7} is an empirical expression that fits all
simulations. Over the range between $10^{-3}<M/M_{\rm nl}<10^2$ it is given by 
\begin{equation}
b_0(x=M/M_{\rm nl})=0.53+0.39x^{0.45}+{0.13 \over 40x+1}+5\times 10^{-4}x^{1.5}. 
\label{bfit}
\end{equation}
This expression should be accurate to about 3\% or better for this model, 
as suggested from the scatter in figure \ref{fig7}.

In figure \ref{fig8} 
we show the bias as a function of mass for several 
simulations for which we varied one parameter at a time, roughly 
spanning the range of interest from cosmological constraints today.  
We see that there is very little difference in the theoretical 
predictions for halo bias as a function of $M/M_{\rm nl}$, suggesting 
that instead of deriving full expressions, which depend on all cosmological 
parameters, one can simply use a single relation with mass in 
units of nonlinear mass, as in equation \ref{bfit}.  
The deviations from this relation are qualitatively consistent with 
the predictions 
given by \cite{1999MNRAS.308..119S}. We can generalize the results from 
equation \ref{bfit} by linearizing the bias relation in terms of 
cosmological parameters,
\begin{eqnarray}
b(x)&=&b_0(x)+\log_{10}(x)[0.4(\Omega_m-0.3+n_s-1) \nonumber \\
&+&0.3(\sigma_8-0.9+h-0.7)+0.8\alpha_s].
\end{eqnarray}
This correction should be reasonable for $1>x>0.1$, while below that 
the correction appears to saturate at 
$-0.4[\Omega_m-0.3+n_s-1]-0.3[\sigma_8-0.9+h-0.7]-0.8 \alpha_s$. 
For massive halos with $M>M_{\rm nl}$ ($x>1$)
the differences among models in the 
bias predictions from \cite{1999MNRAS.308..119S} become larger, but 
this is difficult to observe in these simulations, where the 
number of such halos is small and the bias measurements have
large shot noise.  
In this regime the analytic predictions 
may be more accurate than equation \ref{bfit}: we do not see much evidence 
against the analytic expressions
from our comparisons (figure \ref{fig7}) and analytic 
expressions can be more easily generalized to more general cosmological models. 
Note however
that the simulations used in this paper improve upon the previous 
generation simulations over this regime as well. 

\section{Conclusions}

In this paper we have addressed the relations between the matter density 
field, 
halos and initial density field, focusing on large scales where these
are often assumed to be proportional to each other. 
We focus on two issues. First, what is the scatter between these 
fields around the average relation? This is expressed here in terms 
of relative scatter between the mode amplitudes, which is 
related to the 
stochasticity parameter $r$, defined as the cross-correlation coefficient 
between the two fields. While the two are related we emphasize that 
even small deviations of $r$ from unity may lead to large relative fluctuations 
between the two fields. 
These are of interest whenever one 
is trying to relate the fields to one another to determine 
their relative amplitudes. One example is the bias determination using the
cross-correlation between the weak lensing
signal (tracing the matter density) and the galaxies. 
Another example is the relative bias determination between two different
galaxy populations, which we propose here as an alternative method 
to determine the galaxy bias, because galaxies in low mass halos 
have a bias of $b\sim 0.7$ independent of their mass. 
In all cases we find the scatter between the fields in
individual modes is significant and 
one cannot assume the fields are simply proportional one to another. 
This scatter, coupled with a small number of modes on large scales, 
makes it difficult to accurately determine the bias (or relative bias) and 
needs to be included in the predictions of how accurately can one 
determine the matter power spectrum with these methods. 

The second goal of this paper was to revisit the
halo bias as a function of halo mass. This relation is a 
fundamental ingredient of any halo model 
\citep[see][ for a recent review]{2002PhR...372....1C}
and plays an important role if one is trying to model 
galaxy clustering by connecting it to the underlying halos. 
The previous generation of simulations 
\citep{1998ApJ...503L...9J,1999MNRAS.308..119S} had a limited dynamical range
and the predictions were not tuned specifically for $\Lambda$CDM models. 
As a result the existing expressions overestimate the bias by as much as 
20\% in the 
range below the nonlinear mass, which is likely to be 
the mass range for halos that host most of the galaxies. We propose a 
new expression that fits the simulations better. 
We argue that this expression should be fairly accurate for other 
cosmological models of interest as well, as long as the mass is
expressed in units of nonlinear mass. We give corrections for small 
deviations from this model. The overall accuracy on bias-halo mass 
relation is at the level of 0.03 or better (for $b<1$), which should 
help with the bias determination from  
the current generation of observations.

We thank  P. McDonald, N. Padmanabhan and L. Teodoro for help and 
useful comments, S. Habib
	for encouraging this collaboration and Zheng Zheng and 
David Weinberg for comments on the manuscript.
US is supported by Packard Foundation, Sloan Foundation,
NASA NAG5-1993 and NSF CAREER-0132953.
Simulations by MSW were performed on the Space Simulator Beowulf
Cluster and ASCI Q at Los Alamos.  Work by MSW was performed under the auspices of the
U.S. Dept. of Energy, and supported by its contract \#W-7405-ENG-36 to
Los Alamos National Laboratory.

     \bibliography{apjmnemonic,cosmo,cosmo_preprints}
\bibliographystyle{mnras}

\end{document}